\documentclass[11pt]{article}
\usepackage[margin=1in]{geometry}
\usepackage{amsmath,amssymb,amsthm,mathtools}
\usepackage{graphicx,booktabs}
\usepackage[authoryear,round]{natbib}
\usepackage[colorlinks,citecolor=blue,urlcolor=blue,linkcolor=blue]{hyperref}
\usepackage{microtype}
\theoremstyle{plain}\newtheorem{proposition}{Proposition}\newtheorem{corollary}[proposition]{Corollary}
\theoremstyle{definition}\newtheorem{definition}[proposition]{Definition}
\newcommand{\dTV}{d_{\mathrm{TV}}}
\newcommand{\Neff}{N_{\mathrm{eff}}}
\title{Where should the null live? Surrogate tests for clustering of threshold-derived air-quality indices, with application to PM$_{2.5}$ exceedances in Chile}
\author{Mauricio Herrera-Mar\'{i}n\\[2pt]
\small Faculty of Engineering, Universidad del Desarrollo, Santiago, Chile\\
\small \texttt{mherrera@udd.cl} \quad ORCID 0000-0002-9604-3077}
\date{Version 2}
\begin{document}
\maketitle
\begin{abstract}
\noindent Air-quality and climate monitoring routinely convert native-resolution measurements into exceedance counts before testing whether high values cluster in time. A surrogate null can be imposed on the derived index or on the native series and then propagated through the same thresholding and aggregation. We show that the two placements can encode different scientific hypotheses and lead to opposite conclusions. In 31 Chilean PM$_{2.5}$ monitoring stations (2018--2025), weekly clustering of days above the 50~$\mu$g~m$^{-3}$ national 24-hour standard was significant at all stations under an index-level IAAFT null but at only two under a daily native-resolution null; a native generative null with seasonal marginals and short-memory Gaussian-copula dependence agreed with the native surrogate at 30 of 31 stations. Across thresholds spanning the WHO guideline (15~$\mu$g~m$^{-3}$) to the Chilean episode-alert level (80~$\mu$g~m$^{-3}$), index-only rejections increased from 8 of 31 to all eligible stations as monthly event probabilities became more seasonally concentrated, and equalizing monthly event probabilities at the 50~$\mu$g~m$^{-3}$ level reduced index-only decisions from 29 to five. We formalize the issue as null transport through a lossy observation map: an exact aggregate-level equivalent of a native null exists if and only if the randomization kernel is lumpable with respect to that map. A within-fiber diagnostic, which perturbs native amplitudes while keeping the observed index exactly fixed, shows that this condition fails in the monitoring data: the native null distribution moved by a median of 18 times its Monte Carlo variability. Weekly clustering of PM$_{2.5}$ exceedances in Chile is therefore fully compatible with seasonality plus short-memory daily persistence, and index-level surrogate rejections should not be read as evidence about daily concentration dynamics.
\end{abstract}


\section{Introduction}\label{sec:intro}

Environmental monitoring rarely leads directly from native-resolution measurements to a statistical test. Daily concentrations, temperatures or flows are converted into threshold exceedances, counts, spell durations or seasonal summaries, and persistence or clustering is then assessed on the derived index. Examples include heat-wave and frost-day indices, dry-spell and low-flow statistics, and pollution-exceedance burdens \citep{Sillmann2013,YangXu2017,BrunnerVoigt2024}. Such indices are meaningful to regulators and to the public, but they discard information. A null hypothesis imposed after that loss need not represent the same hypothesis as a null defined for the process that generated the index.

Surrogate-data tests make this issue concrete because the null hypothesis is defined by what the surrogates preserve. IAAFT surrogates, for example, preserve the empirical marginal distribution exactly and the Fourier amplitudes approximately, and randomize everything else \citep{Theiler1992,SchreiberSchmitz1996,Schreiber1998,SchreiberSchmitz2000}. The surrogate literature has emphasized that such algorithms must be checked against the null they are meant to represent, and that spectral and marginal matching need not preserve higher-order or event-localization structure \citep{Kugiumtzis1999,Venema2006,Keylock2012}. Temporal aggregation can also change stochastic model form even for linear processes \citep{StramWei1986}, and lossy transformations can create or obscure apparent memory \citep{Maraun2004,Ohanissian2008}. Here we address a complementary and practical question: \emph{when a clustering test is applied to a threshold-derived index, does it matter whether the null is placed on the index or on the native series?}

Our motivating application is fine particulate matter in Chile. Daily PM$_{2.5}$ in central and southern Chilean cities shows a pronounced winter maximum, with recurrent multi-day episodes during which concentrations exceed the 24-hour standard of 50~$\mu$g~m$^{-3}$ \citep{MMA2011}; the episodes are the basis of local episode-management systems with alert, pre-emergency and emergency levels. Exceedance counts are a natural summary for such systems, and a natural statistical question is whether high-burden weeks cluster more than expected. We show that the answer depends almost entirely on where the null is placed. In 31 monitoring stations, an index-level IAAFT null declares weekly clustering significant everywhere, whereas a native null applied to daily PM$_{2.5}$ and propagated through the same threshold and weekly aggregation rejects at two stations. A third, generative native null---seasonal monthly marginals with short-memory Gaussian-copula dependence---agrees with the native surrogate.

The paper makes four contributions aimed at applied environmental statisticians.
\begin{enumerate}
\item \emph{A substantive application.} We analyse a prospectively frozen cohort of 31 SINCA stations with three null placements, five policy-relevant thresholds from the WHO guideline to the Chilean episode-alert level, and an intervention that equalizes monthly event probabilities. The results imply that weekly exceedance clustering in Chile is compatible with seasonality plus short-memory daily persistence.
\item \emph{A threshold dose-response.} As the threshold rises and events concentrate in fewer months, the index-level null departs progressively from both native nulls, while the two native nulls remain in agreement. This links the statistical discrepancy to an interpretable environmental quantity, the seasonal spread of event probability.
\item \emph{A criterion and a diagnostic.} We formalize the problem as \emph{null transport} through a lossy observation map and give a lumpability criterion for when an aggregate-level null can reproduce a native one. We then turn the criterion into a practical \emph{within-fiber diagnostic} that can be run on any monitoring series, and apply it to the 31 stations.
\item \emph{Guidance.} We give a short checklist for analysts who test persistence or clustering on derived indices.
\end{enumerate}
Controlled simulations, Gaussian controls and a synthetic within-fiber experiment from a prespecified benchmark are summarized in Section~\ref{sec:sim} and detailed in the Appendix.

\section{Data and exceedance index}\label{sec:data}

\subsection{SINCA PM$_{2.5}$ cohort}
\begin{figure}[t]
\centering
\includegraphics[width=\textwidth]{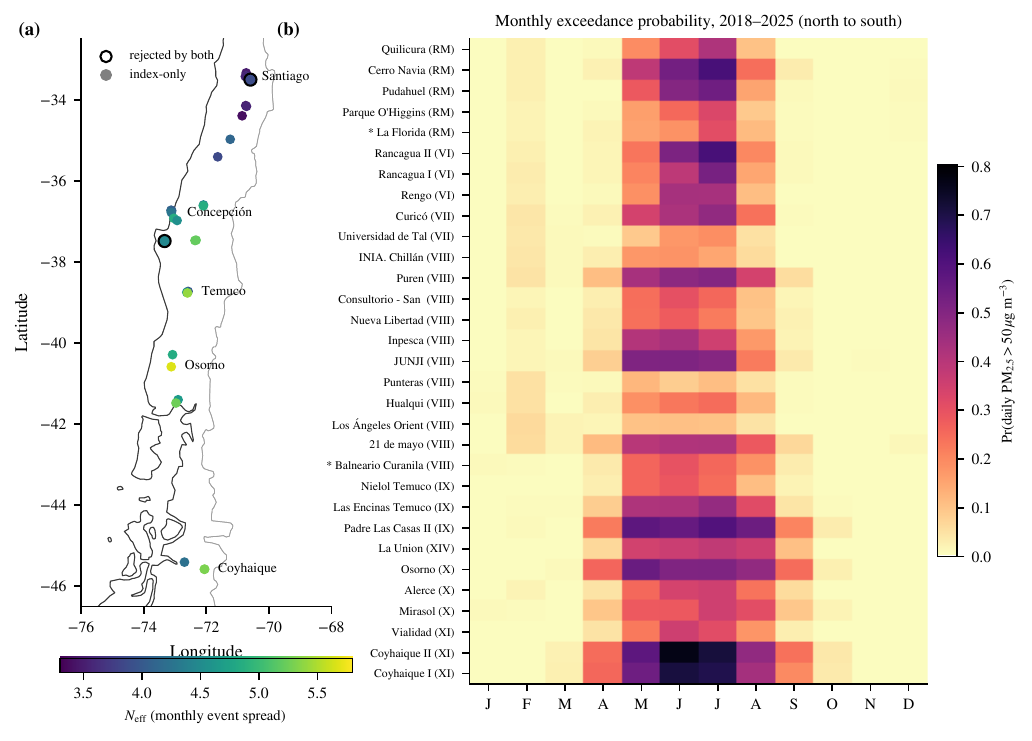}
\caption{SINCA PM$_{2.5}$ cohort. (a) Thirty-one monitoring stations coloured by $\Neff$, the effective number of months with exceedances at 50~$\mu$g~m$^{-3}$; outlined circles mark the two stations at which both the native and the index-level nulls reject. (b) Monthly probability of a day above 50~$\mu$g~m$^{-3}$, 2018--2025, stations ordered from north to south (asterisks mark the same two stations).}
\label{fig:setting}
\end{figure}

Daily PM$_{2.5}$ concentrations were obtained from Chile's National Air Quality Information System (SINCA) through the AtmChile interface \citep{Catalan2022,SINCA2026}. The analysis window runs from 1 January 2018 to 28 December 2025 (417 complete Monday--Sunday weeks). The station cohort was frozen before any surrogate result was computed. Eligibility required an unambiguous station identifier, at least 95\% daily coverage, at least 95\% of weeks with six or more observed days, at least 100 days above 50~$\mu$g~m$^{-3}$, at most 0.1\% of values outside $[0,500]$, and no gross scale-shift flag. Thirty-one stations in eight administrative regions qualified, from Santiago (33.3$^\circ$S) to Coyhaique (45.6$^\circ$S) (Figure~\ref{fig:setting}a).

Exceedances are strongly seasonal at every station (Figure~\ref{fig:setting}b). The probability that a day exceeds 50~$\mu$g~m$^{-3}$ is close to zero from October to March and reaches 0.5--0.75 in June--July at the most affected southern stations (Coyhaique, Osorno, Padre Las Casas). We summarize the seasonal spread of events by
\[
\Neff=\frac{\left(\sum_{m=1}^{12}p_m\right)^2}{\sum_{m=1}^{12}p_m^2},
\]
where $p_m$ is the probability of an event in calendar month $m$; $\Neff=12$ for uniform monthly probabilities and decreases as events concentrate in fewer months. At the 50~$\mu$g~m$^{-3}$ threshold, $\Neff$ ranges from 3.40 to 5.67 (median 4.30): exceedances effectively occupy four to six months a year. Stations further south have larger $\Neff$ (Spearman correlation with latitude $-0.80$) because their high season is longer.

\subsection{Event thresholds and weekly burden}
The primary event is $I_{s,t}=\mathbf 1\{X_{s,t}>50\}$, where $X_{s,t}$ is daily PM$_{2.5}$ at station $s$ and 50~$\mu$g~m$^{-3}$ is the level of the Chilean 24-hour primary standard (D.S.~12/2011). Compliance is assessed through an annual 98th percentile, so a single day above 50 is a policy-relevant event rather than a legal violation. To study threshold dependence we also use the WHO 2021 24-hour guideline (15~$\mu$g~m$^{-3}$) and interim targets 4 and 3 (25 and 37.5~$\mu$g~m$^{-3}$) \citep{WHO2021}, and the lower limit of the Chilean ``alert'' episode level (80~$\mu$g~m$^{-3}$). The threshold sweep was not part of the frozen protocol and is reported as an extension (Section~\ref{sec:status}).

\begin{figure}[t]
\centering
\includegraphics[width=\textwidth]{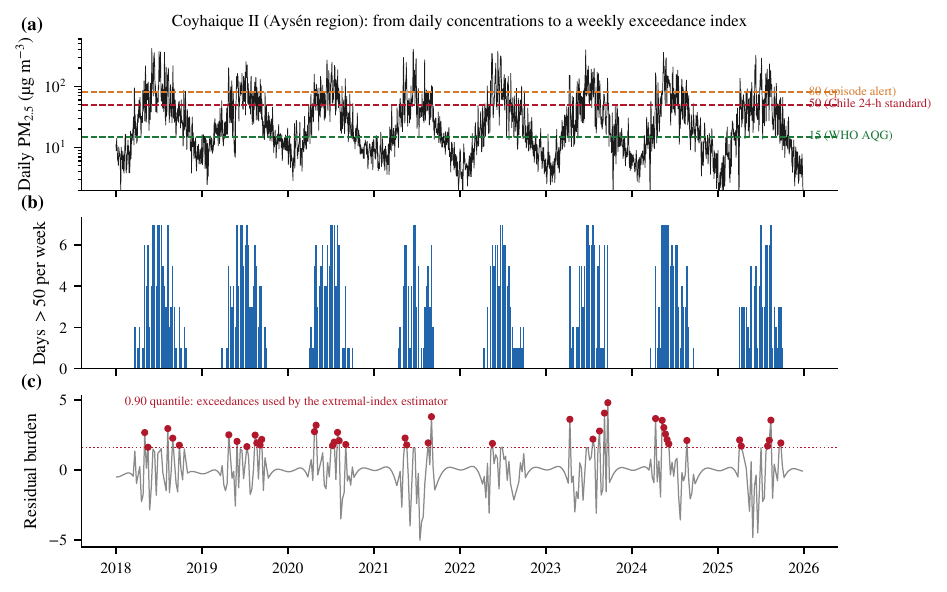}
\caption{From daily concentrations to a weekly exceedance index at Coyhaique II. (a) Daily PM$_{2.5}$ (log scale) with the WHO guideline, the Chilean 24-hour standard and the episode-alert level. (b) Weekly number of days above 50~$\mu$g~m$^{-3}$. (c) Residualized weekly burden; red points are the exceedances of the 0.90 quantile used by the extremal-index estimator.}
\label{fig:pipeline}
\end{figure}

Missing days contribute their station-month event probability. The weekly burden $C_{s,w}=\sum_{t\in w}\widetilde I_{s,t}$ is residualized on a cubic trend and three annual harmonics, and the clustering statistic is the intervals estimator of the extremal index \citep{FerroSegers2003} at the 0.90 quantile of the residual series; smaller values indicate stronger clustering of high-burden weeks. Figure~\ref{fig:pipeline} shows the pipeline for Coyhaique II.

\section{Three null placements}\label{sec:nulls}
All tests are lower-tailed Monte Carlo tests with $B=499$ replicates, $p=\{1+\#(\widehat\theta^\star_b\le\widehat\theta_{\rm obs})\}/(B+1)$, and rejection at $p<0.05$.

\paragraph{Index-level IAAFT.} Thirty-iteration IAAFT is applied directly to the observed weekly residual series. It preserves the weekly empirical marginal exactly and the weekly Fourier amplitudes approximately. The hypothesis is that the \emph{weekly index} is a stationary linear Gaussian process observed through a monotone transformation.

\paragraph{Native conditional surrogate.} Daily PM$_{2.5}$ is standardized by calendar month. Surrogates iteratively match the Fourier amplitudes of the standardized daily series and are rank-remapped within calendar month to the exact observed value multiset (12 iterations; missing days keep their mask). Each surrogate is then thresholded, aggregated, residualized and scored exactly as the data. This null preserves the exact number of exceedances in every calendar month and tests temporal organization beyond that conditioning.

\paragraph{Native generative short-memory null (extension).} As an interpretable reference that does not condition on the observed monthly counts, daily values are converted to normal scores using within-month empirical distribution functions, an autoregression of order $p\le14$ is fitted by Yule--Walker equations with order selected by AIC, Gaussian AR trajectories are simulated, and values are mapped back through the monthly empirical quantile functions before entering the same pipeline. The hypothesis is that daily PM$_{2.5}$ has seasonal monthly marginals and short-memory, Gaussian-copula persistence. Selected orders ranged from 3 to 14 (median 10), consistent with synoptic-scale persistence of several days to two weeks.

\paragraph{Station-level and network-level summaries.} Each station gives a pair of decisions (native, index). Within each contract, $p$-values are adjusted across stations by the Benjamini--Yekutieli procedure \citep{BenjaminiYekutieli2001}. Because stations within a region are dependent, the prespecified network-level test is an exact one-sided sign-flip test over the eight regions applied to $D_r=(\#\text{index-only})_r-(\#\text{native-only})_r$; its smallest attainable $p$-value is $2^{-8}=0.0039$.

\section{Null transport through coarse-graining}\label{sec:theory}

Let $X$ be the native trajectory, $\mathcal G$ the deterministic preprocessing that produces the derived observation $Y=\mathcal G(X)$ (here thresholding, weekly aggregation and residualization), and $K_X(x,\cdot)$ the native randomization kernel (for example, the native surrogate algorithm applied to observed series $x$). The null that the native hypothesis implies for the index is the pushforward
\begin{equation}
L_x=\mathcal G_\#K_X(x,\cdot),\qquad L_x(B)=K_X\{x,\mathcal G^{-1}(B)\}.
\label{eq:transport}
\end{equation}
An index-level procedure can depend on the data only through $y=\mathcal G(x)$.

\begin{definition}[$\mathcal G$-lumpability]
$K_X$ is $\mathcal G$-lumpable if $L_x=L_{x'}$ whenever $\mathcal G(x)=\mathcal G(x')$.
\end{definition}

\begin{proposition}[Quotient null]\label{prop:quotient}
$K_X$ is $\mathcal G$-lumpable if and only if there is a Markov kernel $\overline K_Y$ on the index space with $\mathcal G_\#K_X(x,\cdot)=\overline K_Y(\mathcal Gx,\cdot)$ for every $x$. On the image of $\mathcal G$, $\overline K_Y$ is unique.
\end{proposition}

When lumpability fails, define for each index value $y$ the fiber $\mathcal F_y=\{x:\mathcal Gx=y\}$ and the fiber diameter $\Delta_L(y)=\sup_{x,x'\in\mathcal F_y}\dTV(L_x,L_{x'})$.

\begin{proposition}[Unavoidable error]\label{prop:halfdiam}
For any index-level kernel $Q_Y$, $\sup_{x\in\mathcal F_y}\dTV\{L_x,Q_Y(y,\cdot)\}\ge\Delta_L(y)/2$.
\end{proposition}

\begin{corollary}[Within-fiber witness]\label{cor:witness}
If $\mathcal Gx=\mathcal Gx'$ but some statistic $T$ satisfies $T_\#L_x\neq T_\#L_{x'}$, then $K_X$ is not $\mathcal G$-lumpable, and no index-level procedure reproduces the native null for both $x$ and $x'$.
\end{corollary}

These results adapt the classical notion of lumpability for Markov chains \citep{BurkeRosenblatt1958,KemenySnell1976,RogersPitman1981,Buchholz1994,Geiger2026} to a randomization kernel on trajectories; proofs and a finite counterexample are in the Appendix. Two points matter for practice. First, Proposition~\ref{prop:quotient} is an existence statement: even if a quotient null exists, a convenient index-level surrogate need not be it. Second, Corollary~\ref{cor:witness} is \emph{testable}. One can construct a partner series $\tilde x$ that leaves the index unchanged and compare the native null distributions from $x$ and $\tilde x$ with the Monte Carlo variability of two independent native ensembles from $x$.

\paragraph{Within-fiber diagnostic.} For threshold-and-aggregate indices, a partner in the same fiber is obtained by permuting the observed daily values within each class defined by calendar month and exceedance status. The exceedance chronology, the weekly index, the residual series and the monthly value multisets are then identical, but the arrangement of concentrations within each class---and hence the daily spectrum on which the native surrogate conditions---changes. We measure the shift by the Wasserstein distance between the two native null distributions of $\widehat\theta$, divided by the distance between two independent native ensembles for the same $x$. A ratio near one indicates that the native null is insensitive to information the index discards; large ratios show that it is not, so that index-level inference cannot reproduce the native hypothesis.

Finally, for a continuous monthly marginal $F_m$ the event $X>u_m$ is equivalent to $F_m(X)>1-p_m$, so threshold geometry enters only through the vector $\mathbf p=(p_1,\ldots,p_{12})$ and the native dependence; $\Neff$ is a convenient one-dimensional summary of $\mathbf p$.

\section{Evidence from controlled simulations}\label{sec:sim}
A prespecified benchmark crossed three monthly short-memory generators (Gaussian AR(1), AR(1) with periodic innovation scale, AR(1) with Student-$t_5$ innovations), three persistence levels $\phi\in\{0.2,0.5,0.8\}$ and five seasonal event profiles with $\Neff\approx12$, 11.2, 9.6, 7.0 and 5.0 at a fixed expected count of three events per year, for 13,500 trajectories ($B=249$). The index-level and native contracts were defined as in Section~\ref{sec:nulls} at annual resolution. The index contract rejected in 12.6\% of trajectories and the native contract in 4.7\%; discordant decisions were 1,084 index-only against 16 native-only (Figure~\ref{fig:S1}). Index-only discordance rose from 4.4\% at $\Neff=12$ to 16.2\% at $\Neff\approx5$, in the same direction in all nine generator-by-persistence strata. In an independently seeded equalization experiment (1,350 trajectories), index-only discordance was 15.9\% under concentrated seasonality and 3.9\% under uniform monthly probabilities.

Gaussian controls located the discrepancy in the observation operator. For a continuous annual mean, a generative oracle, the native contract and the index contract all rejected close to the nominal rate; as the observable moved along a soft-threshold path toward a hard count, the index contract separated progressively (23.3\% rejection against 5.3\% for the oracle and 3.7\% for the native contract in the concentrated hard-count case). A synthetic within-fiber experiment showed native null distributions differing by a median 3.3 times their same-state Monte Carlo distance. These analyses are summarized in the Appendix.

\section{Results for Chilean PM$_{2.5}$}\label{sec:results}

\subsection{Primary analysis: index-level and native nulls disagree}
At 50~$\mu$g~m$^{-3}$, the index-level IAAFT null rejected at all 31 stations and the native conditional null at two (La Florida in Santiago and Balneario Curanilahue in the Biob\'io region; outlined in Figure~\ref{fig:setting}); the paired outcomes were 29 index-only, zero native-only and two joint rejections. The regional sign-flip test gave $p=0.0039$, the smallest attainable value. After Benjamini--Yekutieli adjustment, 27 index-level results and no native result remained significant.

\begin{figure}[t]
\centering
\includegraphics[width=\textwidth]{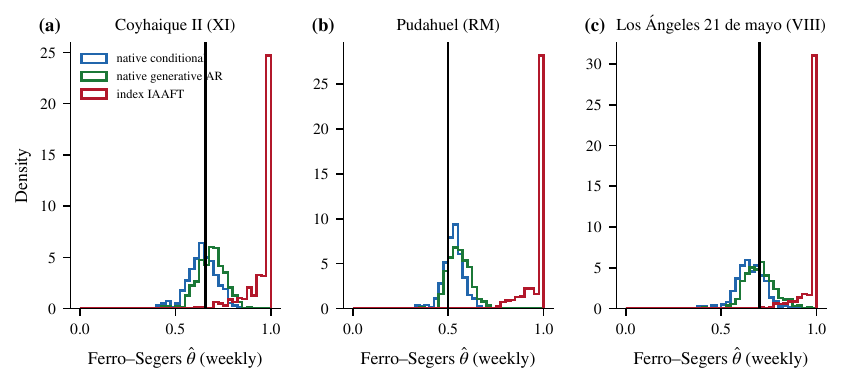}
\caption{Null distributions of the weekly extremal-index estimate at three stations for the 50~$\mu$g~m$^{-3}$ event: native conditional surrogate (blue), native generative short-memory null (green) and index-level IAAFT (red). Vertical lines: observed estimates.}
\label{fig:nulls}
\end{figure}

Figure~\ref{fig:nulls} shows why. The index-level null places most of its mass at or near $\widehat\theta=1$, the value for unclustered exceedances: rearranging a strongly seasonal weekly residual series while matching only its spectrum and marginal tends to scatter the top 10\% of weeks. Both native nulls are centred close to the observed value. Across stations the median observed estimate was 0.535, the median native null centre 0.557 and the median index null centre 0.993, so the median index-minus-native gap was 0.42; by contrast the median ratio of index to native null standard deviation was only 1.12. The discrepancy is therefore a displacement of the null distribution, not a narrowing. A prespecified prediction that seasonal concentration would be associated with narrower index nulls was not supported (Spearman $\rho=-0.38$ between $\Neff$ and the log width ratio; region-cluster bootstrap 95\% interval $-0.62$ to 0.14).

The discrepancy depends on the statistic. With the extremal index at the 0.85 quantile, the index-level null again rejected at all 31 stations and the native null at 10 (21 index-only, none native-only). With a successive-difference energy statistic, which is sensitive to week-to-week roughness rather than to clustering of the highest weeks, the index-level null rejected at four stations and the native null at one. Index-level surrogates are thus particularly misleading for statistics that depend on how extreme weeks are arranged.

\subsection{A generative native null agrees with the native surrogate}
The generative short-memory null rejected at three stations (the two native rejections and Vialidad, Coyhaique). Its decisions agreed with the native conditional surrogate at 30 of 31 stations, the Spearman correlation between the two sets of $p$-values was 0.92, and the median difference between their null centres was 0.023 (Figure~\ref{fig:nulls}). Because this null does not condition on the observed monthly exceedance counts and makes an explicit, interpretable assumption---seasonal marginals and AR-type persistence of normal scores---its agreement indicates that the native conclusion is not an artefact of the constrained surrogate algorithm. Substantively, the weekly clustering of exceedance burden is what one expects from a seasonal concentration regime with persistence over days to two weeks; it does not require additional temporal organization at the weekly scale.

\subsection{Threshold dose-response}
\begin{figure}[t]
\centering
\includegraphics[width=\textwidth]{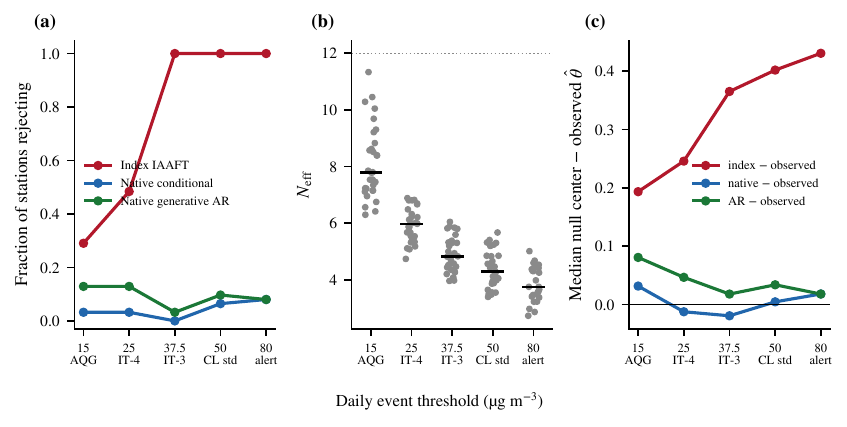}
\caption{Threshold dose-response across WHO and Chilean policy levels (AQG: WHO air-quality guideline; IT: interim target; CL std: Chilean 24-hour standard; alert: lower limit of the Chilean episode-alert level). (a) Fraction of stations rejecting under each null (31 stations, 25 at 80~$\mu$g~m$^{-3}$ where at least 30 events were required). (b) Station $\Neff$; bars mark medians and the dotted line the uniform value 12. (c) Median displacement of each null centre from the observed estimate.}
\label{fig:dose}
\end{figure}

Figure~\ref{fig:dose} repeats the analysis at five thresholds. As the threshold increases from 15 to 80~$\mu$g~m$^{-3}$, events concentrate in fewer months (median $\Neff$ 7.8, 6.0, 4.8, 4.3 and 3.8), and the fraction of stations at which the index-level null rejects rises from 9 of 31 to 31 of 31 at 37.5 and 50~$\mu$g~m$^{-3}$ and to all 25 eligible stations at 80~$\mu$g~m$^{-3}$. The two native nulls reject at between zero and four stations at every threshold, with 26--30 agreeing decisions. The index-minus-observed displacement of the null centre increases monotonically with the threshold, while the native displacements stay close to zero. Index-only decisions were 8, 14, 31, 29 and 23 of the eligible stations, and native-only decisions were zero at every threshold.

The dose-response is consistent with the threshold-geometry argument of Section~\ref{sec:theory}: higher thresholds produce more seasonally concentrated events, and the index-level null departs further from both native nulls. It also shows that the conclusion is not specific to the national standard. For indices built at the WHO guideline level the two placements disagree at about a quarter of stations; for indices relevant to episode management they disagree almost everywhere.

\subsection{Equalizing monthly event probabilities}
A prespecified intervention replaced the fixed threshold by month-specific quantiles that make monthly event probabilities approximately equal while preserving each station's mean event rate. Realized $\Neff$ moved close to 12 at every station. Index-level rejections fell from 31 to six, native rejections from two to one, and index-only decisions from 29 to five; the regional sign-flip $p$-value weakened to 0.0625 and no station remained significant after adjustment. The median index-minus-native null-centre gap fell from 0.42 to 0.10 and decreased at all 31 stations (Figure~\ref{fig:S2}). The intervention is not a policy alternative---equalized thresholds have no regulatory meaning---but it isolates seasonal event geometry as the main driver of the discrepancy.

\subsection{Within-fiber diagnostic in the monitoring data}
\begin{figure}[t]
\centering
\includegraphics[width=\textwidth]{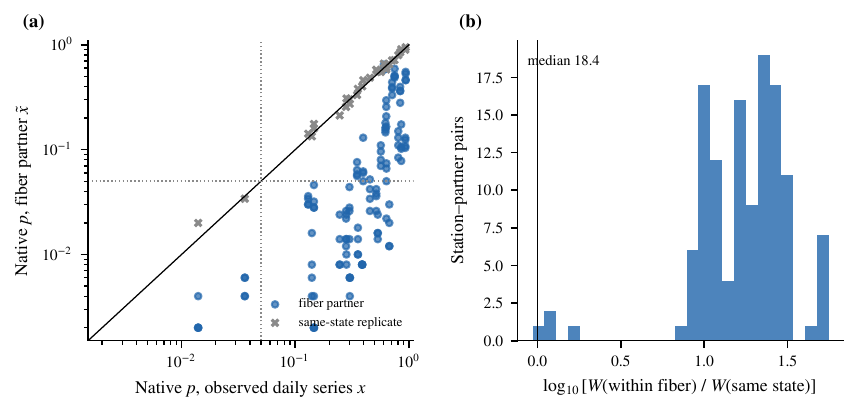}
\caption{Within-fiber diagnostic for the 31 SINCA stations at 50~$\mu$g~m$^{-3}$. (a) Native $p$-values for the observed daily series against those for four fiber partners per station (circles) and for an independent replicate of the observed-series null (crosses); dotted lines mark 0.05. (b) Ratio of the Wasserstein distance between observed and partner native nulls to the distance between two independent native nulls for the observed series.}
\label{fig:fiber}
\end{figure}

For each station we generated four fiber partners at 50~$\mu$g~m$^{-3}$, each with exactly the observed exceedance chronology and weekly index, and computed native null distributions for the observed series, an independent replicate, and each partner (Figure~\ref{fig:fiber}). The replicate reproduced the observed native $p$-values closely, but the partners did not. The median ratio of within-fiber to same-state Wasserstein distance was 18.4 (124 pairs), the partner null centre moved upward by a median 0.094 (towards less clustering) and its spread fell slightly (median standard-deviation ratio 0.91). The median native $p$-value was 0.45 for the observed series and 0.036 for its partners; 55\% of partner tests rejected, and 18 of 31 stations had at least one rejecting partner, against 2 of 31 for the observed series.

By Corollary~\ref{cor:witness}, the native null is not lumpable with respect to the threshold-and-aggregation map in these data, so no index-level surrogate can reproduce it. The direction of the change is also informative. Permuting values within month-by-exceedance classes destroys the day-to-day persistence of concentrations \emph{inside} episodes and quiet spells while leaving the exceedance pattern intact. The native null then generates less clustered weekly burdens, and the observed clustering becomes significant. In other words, the native null ``explains'' weekly clustering through the persistence of daily concentrations, which is exactly the information that thresholding and aggregation discard and that an index-level surrogate cannot see.

\subsection{Sensitivity}
An exposure-standardized weekly count, defined at 22 stations, gave 22 index-level and one native rejection. Increasing IAAFT iterations from 30 to 100 or 300 in the simulations did not remove the discrepancy, and a phase-dispersion diagnostic did not support a phase-locking explanation (Appendix).

\section{Discussion}\label{sec:discussion}

\paragraph{What the application shows.} Placing the null on the weekly index leads to the conclusion that high-burden weeks cluster unusually at every Chilean station. Placing it on daily PM$_{2.5}$---either with a constrained surrogate or with an explicit seasonal short-memory model---leads to the conclusion that the observed clustering is what seasonality and day-to-day persistence produce. The two answers address different hypotheses. The index-level test asks whether the weekly series is consistent with a linear Gaussian process with the same spectrum, which a strongly seasonal, threshold-generated series is not; it rejects largely because of the shape of the index. The native tests ask whether daily concentrations contain temporal organization beyond seasonal marginals and short memory, and they find little evidence that they do.

For air-quality management this distinction matters. Episode-management systems and epidemiological studies often reason about runs of high-pollution days. Our results suggest that, at weekly resolution, such runs in Chile are adequately described by seasonal concentration regimes and synoptic-scale persistence, so forecasting and alert systems can rely on models of daily persistence rather than on an additional weekly clustering mechanism. Conversely, an analyst who reported the index-level result as evidence of ``anomalous clustering of pollution episodes'' would be drawing a conclusion about daily dynamics that the test cannot support.

\paragraph{A general mechanism.} The threshold dose-response and the equalization intervention identify the main driver: the more concentrated the events are in a few months, the more the index-level null departs from any native null. This mechanism is not specific to PM$_{2.5}$. Heat-wave counts, frost days, dry spells and low-flow indices are all threshold-derived and often strongly seasonal \citep{Sillmann2013,YangXu2017,BrunnerVoigt2024}, and the same issue arises with permutation, block-bootstrap and other resampling nulls whenever they are placed after a lossy transformation \citep{DavisonHinkley1997}. The within-fiber diagnostic requires only the native series, the index pipeline and the chosen native null, and it can be run before interpreting any index-level clustering test.

\paragraph{Recommendations.} We suggest four checks. (i) State whether the scientific hypothesis concerns the native process or only the derived index. (ii) State what the surrogate or resampling procedure conditions on, and at what resolution. (iii) When a native interpretation is intended, apply the null to the native series and transport it through the index pipeline; use the within-fiber diagnostic to check whether information discarded by the index matters to the null. (iv) For threshold-derived indices, report the seasonal spread of event probabilities (for example $\Neff$) and repeat the analysis at more than one threshold.

\paragraph{Limitations.} The magnitudes reported are specific to the surrogate constructions, the extremal-index statistic and the weekly aggregation studied. The native conditional surrogate is highly constrained and has limited power for some alternatives (Appendix); native non-rejection is therefore not evidence that daily PM$_{2.5}$ has no structure beyond short memory, although the agreement with the generative null strengthens the interpretation. The generative null assumes a Gaussian copula and stationary AR dependence within the analysis window and does not model meteorological covariates. The analytical counterexample in the Supplement establishes structural possibility for a simplified kernel rather than a theorem about IAAFT. The SINCA cohort and primary protocol were frozen before any application-level surrogate output was inspected, but the threshold sweep, the generative null and the real-data within-fiber diagnostic were added afterwards and should be read as extensions. Finally, the station network is not a probability sample of population exposure, and a day above 50~$\mu$g~m$^{-3}$ is not an individual legal violation.

\section{Conclusions}\label{sec:conclusion}
Coarse-graining changes not only the data available for inference but also the hypothesis a resampling null represents. For threshold-derived environmental indices, a null placed on the index can be internally coherent and still be the wrong null for a question about the native process. In Chilean PM$_{2.5}$ monitoring this choice reverses the conclusion about weekly exceedance clustering at nearly every station; native nulls, including an explicit seasonal short-memory model, indicate that the clustering is explained by seasonality and daily persistence. The lumpability criterion explains when an index-level null can stand in for a native one, and the within-fiber diagnostic shows directly, in the monitoring data, that here it cannot.

\section*{Statement of analysis status}\label{sec:status}
The SINCA cohort and the primary protocol (50~$\mu$g~m$^{-3}$ event, weekly burden, extremal-index statistic, native conditional and index-level IAAFT nulls, Benjamini--Yekutieli adjustment, regional sign-flip test, equalization intervention and width prediction) were frozen before any application-level surrogate result was inspected. The simulation benchmark and the equalization simulation were prespecified; the Gaussian and synthetic within-fiber analyses were designed after the benchmark. The threshold sweep, the generative short-memory null and the within-fiber diagnostic on SINCA data are extensions added for this version and are reported as such.

\section*{Data availability statement}
The reproducibility package (frozen cohort and protocol, analysis-ready 2018--2025 PM$_{2.5}$ extract, scripts, station-level outputs, simulation outputs and diagnostics) is archived on Zenodo, DOI \href{https://doi.org/10.5281/zenodo.22713989}{10.5281/zenodo.22713989}, with the code for the extension analyses added in a new release. Source observations are publicly available from SINCA.

\section*{Funding and conflict of interest}
The author received no external funding and declares no conflict of interest.

\bibliographystyle{plainnat}
\bibliography{references}

\clearpage
\appendix
\renewcommand{\thefigure}{A\arabic{figure}}\setcounter{figure}{0}
\renewcommand{\thetable}{A\arabic{table}}\setcounter{table}{0}
\renewcommand{\thesection}{A\arabic{section}}
\begin{center}\Large\bfseries Appendix: technical derivations, diagnostics and additional analyses\end{center}

\section{Figures referenced in the main text}
\begin{figure}[h]\centering\includegraphics[width=\textwidth]{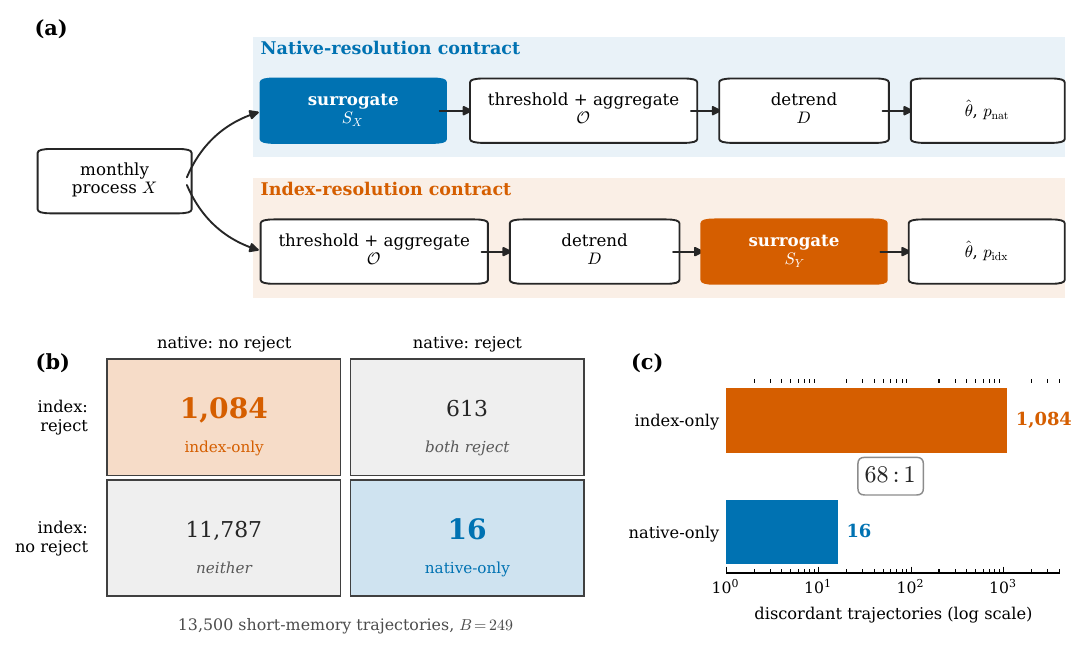}
\caption{Prespecified 13,500-trajectory benchmark: native and index-level contracts act on opposite sides of a threshold-and-aggregation pipeline; 1,084 trajectories rejected only under the index-level contract and 16 only under the native contract.}\label{fig:S1}\end{figure}
\begin{figure}[h]\centering\includegraphics[width=\textwidth]{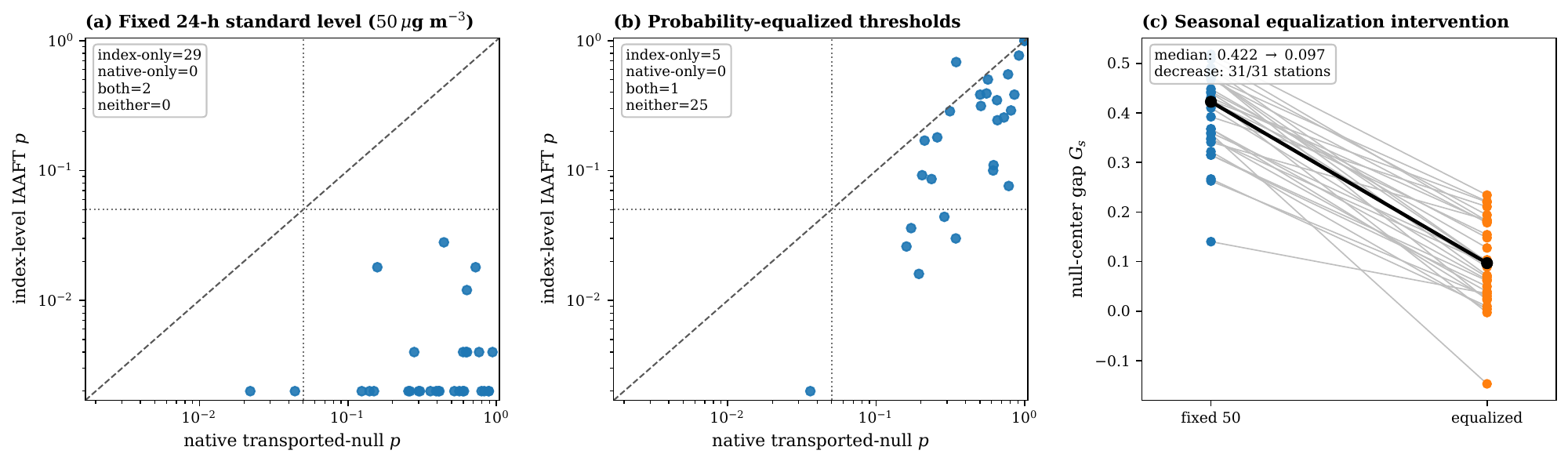}
\caption{SINCA application under the fixed 50 $\mu$g m$^{-3}$ event and under probability equalization: equalization reduces index-only decisions from 29 to five and reduces the index-minus-native null-centre gap at every station.}\label{fig:S2}\end{figure}
\begin{figure}[h]\centering\includegraphics[width=0.72\textwidth]{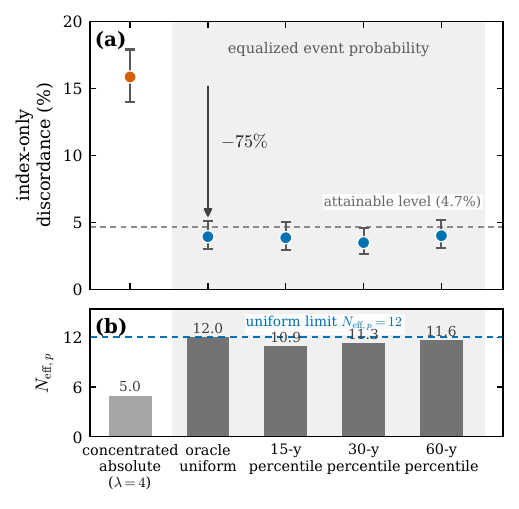}
\caption{Independent 1,350-trajectory equalization experiment: index-only discordance 15.85\% under concentrated seasonality and 3.93\% under uniform monthly probabilities.}\end{figure}
\begin{figure}[h]\centering\includegraphics[width=\textwidth]{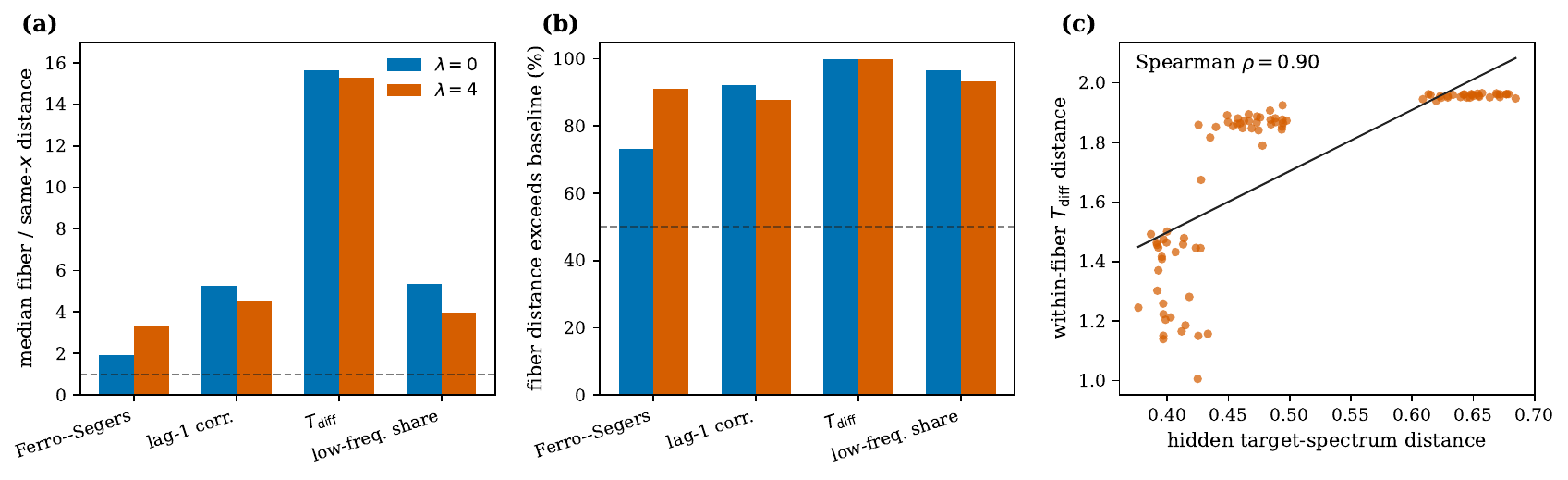}
\caption{Synthetic within-fiber experiment: pairs with identical threshold chronology, aggregate counts and monthly value multisets have native pushed-forward nulls that differ beyond same-state Monte Carlo variability.}\end{figure}
\clearpage
\section{Extension analyses on the SINCA cohort}
\paragraph{Generative short-memory null.} For each station, observed daily values in calendar month $m$ were converted to normal scores $\Phi^{-1}\{r/(n_m+1)\}$, with $r$ the within-month rank. Missing days were set to zero for estimation. Autocovariances of the score series up to lag 14 were computed, AR($p$) models with $1\le p\le14$ were fitted by the Yule--Walker equations, and $p$ was chosen by AIC. Each replicate simulated a Gaussian AR($p$) path (burn-in 200 days), standardized it, applied $\Phi$, and mapped the uniform values through the within-month empirical quantile function of the observed data. Missing days were reinserted with the original mask and then handled exactly as in the primary pipeline. Seeds derive from base 20261007 and the station identifier.
\paragraph{Threshold sweep.} The same three nulls were applied at 15, 25, 37.5, 50 and 80 $\mu$g m$^{-3}$ with $B=499$; Table~\ref{tab:sweep} summarizes the results.
\paragraph{Within-fiber diagnostic.} For each station, two independent native ensembles were generated for the observed series and one for each of four partners obtained by permuting values within calendar-month $\times$ exceedance-status classes (Table~\ref{tab:fiberSINCA}).
\begin{table}[h]\centering\small\caption{Threshold sweep (extension). Counts of stations rejecting at $p<0.05$ ($B=499$) under each null; eligibility required at least 30 events and 30 non-events.}\label{tab:sweep}
\begin{tabular}{lrrrrrrrr}\toprule Threshold ($\mu$g m$^{-3}$) & Stations & median $N_{\rm eff}$ & Index & Native & AR & Index-only & Native-only & AR=Native\\\midrule
15 & 31 & 7.78 & 9 & 1 & 4 & 8 & 0 & 26\\
25 & 31 & 5.98 & 15 & 1 & 4 & 14 & 0 & 28\\
37.5 & 31 & 4.82 & 31 & 0 & 1 & 31 & 0 & 30\\
50 & 31 & 4.30 & 31 & 2 & 3 & 29 & 0 & 30\\
80 & 25 & 3.75 & 25 & 2 & 2 & 23 & 0 & 25\\
\bottomrule\end{tabular}\end{table}
\begin{table}[h]\centering\small\caption{Within-fiber diagnostic by station at 50 $\mu$g m$^{-3}$ (four partners per station). $W$ ratio: Wasserstein distance between observed and partner native nulls divided by the distance between two independent native nulls for the observed series.}\label{tab:fiberSINCA}
\begin{tabular}{llrrrrr}\toprule Station & Region & $p$ (observed) & $p$ (replicate) & min $p$ (partners) & median $p$ (partners) & median $W$ ratio\\\midrule
CHLI & VIII & 0.302 & 0.274 & 0.004 & 0.006 & 48.8\\
CHLII & VIII & 0.602 & 0.662 & 0.578 & 0.592 & 1.2\\
CHYI & VIII & 0.130 & 0.142 & 0.030 & 0.032 & 10.5\\
CN & RM & 0.388 & 0.402 & 0.008 & 0.008 & 53.5\\
CNH & VIII & 0.036 & 0.034 & 0.004 & 0.005 & 9.2\\
COI & XI & 0.712 & 0.710 & 0.332 & 0.389 & 11.4\\
COII & XI & 0.632 & 0.582 & 0.166 & 0.232 & 10.0\\
CRCI & VII & 0.838 & 0.850 & 0.280 & 0.363 & 24.4\\
HPNIII & VIII & 0.936 & 0.948 & 0.460 & 0.496 & 24.3\\
HQI & VIII & 0.350 & 0.332 & 0.056 & 0.061 & 21.2\\
LAI & VIII & 0.748 & 0.716 & 0.488 & 0.505 & 10.8\\
LAII & VIII & 0.396 & 0.460 & 0.050 & 0.061 & 10.2\\
LF & RM & 0.014 & 0.020 & 0.002 & 0.002 & 10.5\\
LU & XIV & 0.454 & 0.486 & 0.026 & 0.038 & 15.5\\
OS & X & 0.566 & 0.550 & 0.070 & 0.088 & 14.1\\
PLCII & IX & 0.856 & 0.908 & 0.078 & 0.104 & 17.3\\
PMI & X & 0.806 & 0.798 & 0.084 & 0.120 & 29.5\\
PMII & X & 0.930 & 0.898 & 0.104 & 0.116 & 26.0\\
PU & RM & 0.280 & 0.256 & 0.012 & 0.018 & 24.4\\
QU & RM & 0.246 & 0.212 & 0.008 & 0.011 & 17.1\\
RGAI & VI & 0.516 & 0.578 & 0.026 & 0.037 & 16.6\\
RGAII & VI & 0.532 & 0.556 & 0.014 & 0.016 & 31.7\\
RNG & VI & 0.300 & 0.302 & 0.014 & 0.027 & 22.4\\
SA & RM & 0.140 & 0.134 & 0.004 & 0.007 & 25.0\\
TALIII & VII & 0.146 & 0.160 & 0.028 & 0.030 & 12.0\\
TE & IX & 0.634 & 0.608 & 0.022 & 0.058 & 21.6\\
TEII & IX & 0.282 & 0.308 & 0.008 & 0.012 & 21.9\\
THNI & VIII & 0.622 & 0.612 & 0.146 & 0.158 & 18.1\\
THNV & VIII & 0.356 & 0.380 & 0.010 & 0.012 & 11.0\\
THNVI & VIII & 0.672 & 0.652 & 0.012 & 0.016 & 33.0\\
VI & XI & 0.146 & 0.176 & 0.002 & 0.002 & 28.4\\
\bottomrule\end{tabular}\end{table}
\clearpage

\section{Status of the analyses}\label{supp:status}

The research program distinguishes three analysis statuses. The main
benchmark of 13,500 trajectories, concentration contrast, probability equalization,
native gate sensitivity study, and grouping checks were specified before the
mechanism interpretation developed. Gaussian controls and the later threshold,
spacing, iteration, phase, and fiber diagnostics were designed after the
confirmatory benchmark. They are treated as mechanism diagnostics.

The SINCA PM$_{2.5}$ analysis occupies a third category. It was designed after
the null-transport framework and the synthetic findings were known, but its
station cohort and full inferential protocol were frozen before any
application-level surrogate output was inspected. It is therefore a
prospectively specified applied validation rather than part of the original
confirmatory simulation or a post hoc mechanism diagnostic. The exact station-list SHA-256 is
\begin{center}
\texttt{59b019cd344b1cdf9c0f6fb90b381ad12}\\
\texttt{d416d72461d94d31d6ce0039ff010e6}
\end{center}
and is recorded in the frozen protocol manifest.

\section{Proofs for the null-transport framework}\label{supp:proofs}

\subsection{Quotient-null criterion}

Let $(\mathsf X,\mathcal A)$ be the native trajectory space,
$(\mathsf Y,\mathcal B)$ the derived-data space,
$\mathcal G:\mathsf X\rightarrow\mathsf Y$ a measurable deterministic map,
and $K_X$ a Markov kernel on $\mathsf X$. For
\[
L_x(B)=K_X\{x,\mathcal G^{-1}(B)\},
\]
strong $\mathcal G$-lumpability means $L_x=L_{x'}$ whenever
$\mathcal G(x)=\mathcal G(x')$.

\begin{proposition}
Equip $\mathsf Y=\mathcal G(\mathsf X)$ with the quotient sigma-algebra
induced by $\mathcal G$. Strong $\mathcal G$-lumpability holds if and only if
there exists a Markov kernel $\overline K_Y$ satisfying
\[
\mathcal G_\#K_X(x,\cdot)=\overline K_Y(\mathcal Gx,\cdot)
\quad\text{for every }x\in\mathsf X.
\]
On $\mathcal G(\mathsf X)$ the kernel is unique.
\end{proposition}

\begin{proof}
If the displayed identity holds and $\mathcal Gx=\mathcal Gx'$, both
transported laws equal $\overline K_Y(\mathcal Gx,\cdot)$, proving
lumpability. Conversely, under lumpability define
$\overline K_Y(y,B)=L_x(B)$ for any
$x\in\mathcal G^{-1}(\{y\})$. The value is independent of the representative
because all native states in a fiber have the same transported law.
Measurability follows from the quotient sigma-algebra. Uniqueness on the
image of $\mathcal G$ is immediate.
\end{proof}

\subsection{Unavoidable error for an aggregate-only null}

For $y\in\mathsf Y$, let
$\mathcal F_y=\{x:\mathcal Gx=y\}$ and
\[
\Delta_L(y)=\sup_{x,x'\in\mathcal F_y}\dTV(L_x,L_{x'}).
\]

\begin{proposition}
For every aggregate-only kernel $Q_Y(y,\cdot)$,
\[
\sup_{x\in\mathcal F_y}\dTV\{L_x,Q_Y(y,\cdot)\}
\ge \frac12\Delta_L(y).
\]
\end{proposition}

\begin{proof}
For any $x,x'\in\mathcal F_y$, the triangle inequality gives
\[
\dTV(L_x,L_{x'})
\le \dTV\{L_x,Q_Y(y,\cdot)\}
+\dTV\{Q_Y(y,\cdot),L_{x'}\}.
\]
At least one term on the right is at least one half of the left side.
Taking the supremum over $x,x'$ gives the result.
\end{proof}

\begin{corollary}
If $\mathcal Gx=\mathcal Gx'$ and there is a measurable statistic $T$
for which $T_\#L_x\ne T_\#L_{x'}$, then $K_X$ is not strongly
$\mathcal G$-lumpable.
\end{corollary}

\begin{proof}
Strong lumpability would imply $L_x=L_{x'}$ and therefore equality after
every measurable pushforward. The contrapositive proves the claim.
\end{proof}

The data-processing inequality also gives
\[
\dTV(T_\#L_x,T_\#L_{x'})
\le \dTV(L_x,L_{x'}),
\]
which motivates separate consideration of discrepancies in the kernel, the
projected statistic, and the final decision.

\subsection{Model-dependent transport}

Failure of strong lumpability does not preclude aggregate-space inference, but
it makes the result model dependent. Let $P_0$ be a benchmark distribution on
$\mathsf X$ with regular conditional distribution
$\mu_y(dx)=P_0(dx\mid \mathcal G(X)=y)$. Define
\[
\overline K_{P_0}(y,B)
=\int_{\mathcal F_y}L_x(B)\,\mu_y(dx).
\]
If $X\sim P_0$ and $X^\star\mid X\sim K_X(X,\cdot)$, then
\[
\Pr\{\mathcal G(X^\star)\in B\mid \mathcal G(X)=y\}
=\overline K_{P_0}(y,B).
\]
Under strong lumpability this reduces to the unique quotient kernel for every
$P_0$. Otherwise, the aggregate alone does not identify the native inferential
contract because unresolved native states must be weighted by an additional
model.

\section{Finite spectral-conditioning counterexample}\label{supp:counterexample}

This example isolates a mechanism relevant to marginal-and-spectrum
surrogates. It is not an analytical representation of IAAFT.

Let
\[
\Omega=\left\{z\in\{0,1\}^{4}:\sum_{t=1}^{4}z_t=2\right\},
\qquad
\mathcal O(z)=(z_1+z_2,z_3+z_4).
\]
With $\bar z=1/2$, define the circular lag-one score
\[
R(z)=\sum_{t=1}^{4}(z_t-\bar z)(z_{t+1}-\bar z),
\qquad z_5=z_1.
\]
By the discrete Wiener--Khinchin identity, $R$ is a linear functional of
Fourier power. For $\tau\ge0$, define the marginal-preserving
spectral-tilted kernel
\[
K_\tau(x,z)
=
\frac{\exp\{-\tau[R(z)-R(x)]^2\}}
{\sum_{w\in\Omega}\exp\{-\tau[R(w)-R(x)]^2\}}.
\]
Every draw has exactly two ones, so the empirical binary marginal is preserved.

Take two native states $x,x'\in\Omega$ with
$\mathcal O(x)=\mathcal O(x')=(1,1)$ but different $R$ values. Enumerating the
six elements of $\Omega$ gives two spectral classes within the same aggregate
fiber. If $q=e^{-\tau}$, direct summation of the pushed-forward probabilities
gives
\[
\dTV(L_x,L_{x'})
=\frac{1-q^2}{2q^2+5q+2},
\qquad \tau>0.
\]
The value is strictly positive for every $\tau>0$. Thus exact empirical
marginal preservation plus microstate spectral conditioning can be
non-lumpable under aggregation. The role of this construction is structural:
the implemented IAAFT/native contracts are assessed by the empirical
within-fiber experiment rather than by identifying them with $K_\tau$.

\section{Threshold geometry and Gaussian covariance}\label{supp:threshold}

Let $F_m$ be a continuous monthly marginal and let
$U_{y,m}=F_m(X_{y,m})$. The threshold event
$X_{y,m}<u_m$ is equivalent to
\[
U_{y,m}<p_m,\qquad p_m=F_m(u_m).
\]
Conditional on the temporal copula, the entire indicator process depends on
the marginal/threshold construction through
$\mathbf p=(p_1,\ldots,p_{12})$. The effective seasonal probability support
\[
\Neff=\frac{(\sum_m p_m)^2}{\sum_m p_m^2}
\]
equals 12 under a uniform profile and decreases as probability mass
concentrates in fewer months.

For the stationary Gaussian AR(1) branch,
$X_t=\phi X_{t-1}+\sqrt{1-\phi^2}\varepsilon_t$, write
$I_{y,m}=\mathbf 1\{X_{12(y-1)+m}<u_m\}$. Then
\[
\operatorname{Cov}(I_{y,m},I_{y+k,n})
=
\Phi_2\{u_m,u_n;\phi^{|12k+n-m|}\}-p_mp_n,
\]
and the annual-count covariance is
\[
\gamma_C(k)=
\sum_{m=1}^{12}\sum_{n=1}^{12}
\left[
\Phi_2\{u_m,u_n;\phi^{|12k+n-m|}\}-p_mp_n
\right].
\]
This formula shows why $\Neff$ alone cannot determine the aggregated
dependence law: the full event-probability vector and the native correlation
structure both enter.

\section{Detailed surrogate contracts}\label{supp:contracts}

\subsection{Index-level IAAFT}

For an observed derived series $z$, an index surrogate begins from a random
permutation. Each iteration replaces the current Fourier amplitudes by those
of the centered observed series while retaining phases, inverse transforms,
and rank-remaps to the exact sorted values of $z$. The normalized spectral
error is
\[
E_{\rm spec}
=
\frac{\operatorname{mean}(|\widehat z^\star|-|\widehat z|)^2}
{\operatorname{mean}|\widehat z|^2+10^{-15}}.
\]
The confirmatory synthetic and SINCA index contracts used at most 30
iterations.

\subsection{Synthetic native contract}

For each calendar month, observed monthly values were standardized by the
month-specific mean and standard deviation. The target Fourier amplitudes were
computed from the full standardized monthly sequence. Surrogates were
initialized by permuting exact observed values within calendar month, projected
onto the target amplitudes, inverse transformed, and rank-remapped within each
calendar-month group to the exact observed value multiset. The primary
contract used 12 iterations. A prespecified robustness contract used calendar
month $\times$ 50-year blocks.

\subsection{SINCA native contract}

The daily SINCA contract used the same logic with a fixed missing mask. For
each station and month, observed values were standardized using observed
month-specific means and standard deviations. Missing positions were set to
zero only in the complete standardized sequence used to define the FFT target.
Surrogate initialization permuted raw observed values among observed positions
within calendar month. At every iteration, observed positions were
rank-remapped back to the exact sorted observed raw values in that month, while
missing positions were reset to the month mean. Exceedances were evaluated
only at originally observed positions; the deterministic missing-day
contributions used in the observed weekly burden were then reinserted before
weekly aggregation. The contract therefore preserved exact observed
calendar-month value multisets and fixed-threshold exceedance counts.

\section{Confirmatory robustness and audit}\label{supp:audit}

The primary benchmark of 13,500 trajectories produced 1697 index-level rejections,
629 native rejections, 1084 index-only decisions, and 16 native-only
decisions. Replacing full-record calendar-month grouping by the prespecified
50-year grouping left the direction unchanged: native rejection was 4.95\%
and index-only discordance 7.86\%.

An independently written implementation selected ten trajectories from each of
the 45 generator-family $\times\phi\times\lambda$ cells, giving 450
trajectories. Deterministic regeneration reproduced observed Ferro--Segers
values to machine precision (maximum absolute discrepancy
$1.11\times10^{-16}$), and independent arithmetic reaggregation recovered the
stored confirmatory decision counts exactly. A new $B=99$ surrogate rerun on
the 450-trajectory subset produced 24 index-only and zero native-only
decisions, reproducing the direction without being intended to reproduce the
full-sample rates.

Clipping the raw Ferro--Segers estimator at one did not create the positive
index/native offset. Under the concentrated profile, the median raw annual
offset was $+0.1261$ before clipping and $+0.0181$ after clipping, so clipping
attenuated rather than generated the difference.

\section{Native-gate sensitivity}\label{supp:power}

The native synthetic gate was evaluated under two injected alternatives while
preserving exact calendar-month value multisets and threshold-event counts.
Persistent-regime alignment gave full-gate detection rates 6.7\%, 9.3\%,
10.7\%, and 8.0\% for $\kappa=0.2,0.4,0.6,0.8$. History feedback gave 6.7\%,
20.7\%, 30.0\%, and 30.7\% for
$\kappa=0.25,0.50,0.75,1.00$. The benchmark-null gate rate was 4.0\%.
These results are why native non-rejection is not interpreted as evidence of
absence of temporal organization.

\begin{figure}[t]
\centering
\includegraphics[width=0.70\textwidth]{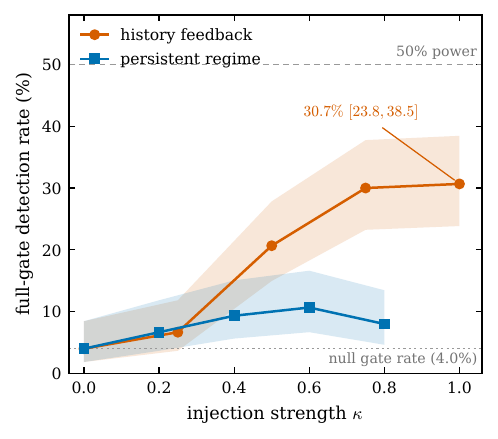}
\caption{Mechanism-specific sensitivity of the native synthetic gate. The
prespecified alternatives have different detectability, and neither grid
reaches 50\% power.}
\label{fig:supppower}
\end{figure}

\section{Gaussian and soft-threshold diagnostics}\label{supp:ablation}

The diagnostic used the stationary Gaussian AR(1) branch at three
persistence values, $\phi=0.2$, $0.5$, and $0.8$. There were 100 independent
trajectories per value and $B=249$ surrogates. Three null ensembles were
compared: independent draws from the known generator, the native constrained
surrogate followed by the chosen operator, and annual IAAFT after the
operator.

For the raw continuous annual Gaussian mean, oracle, native, and index-level
rejection rates were 5.3\%, 3.7\%, and 4.7\%, respectively. After cubic
detrending they were 6.3\%, 4.0\%, and 3.7\%. The large index-only asymmetry
was therefore absent for the continuous control.

For the hard annual count with uniform monthly event probability
($\lambda=0$), oracle and native rejection were both 7.7\% and annual IAAFT
rejection was 11.0\%. Under concentrated event geometry ($\lambda=4$),
oracle rejection was 5.3\%, native rejection 3.7\%, and annual IAAFT rejection
23.3\%; 59 trajectories were index-only and none native-only.

\begin{figure}[t]
\centering
\includegraphics[width=\textwidth]{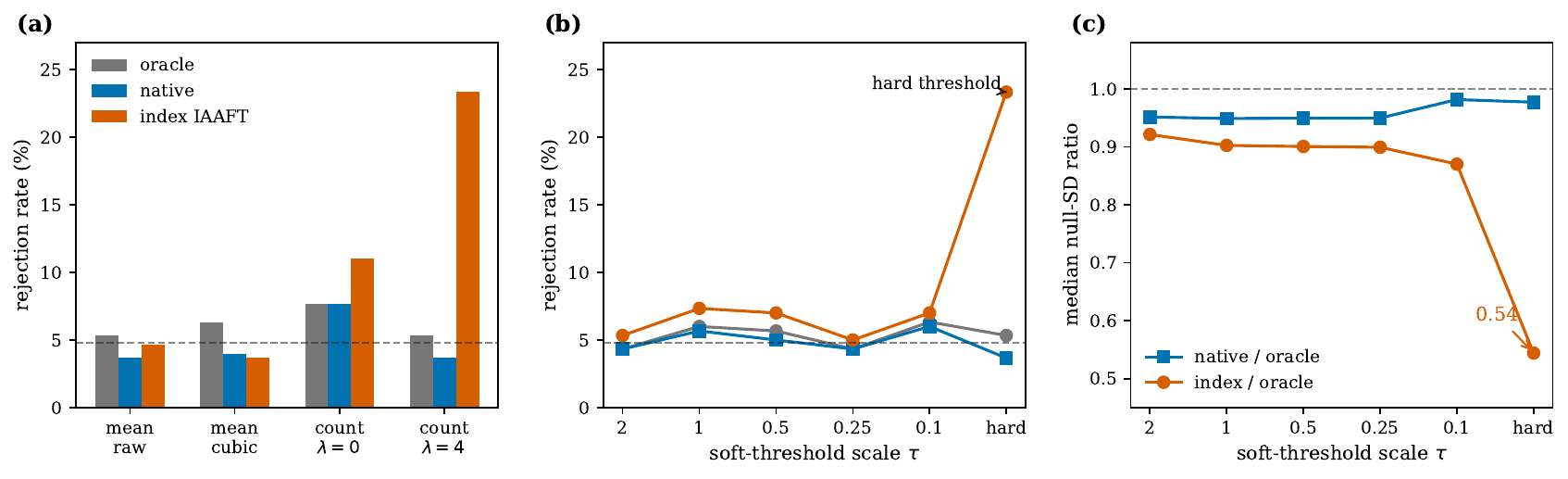}
\caption{Gaussian oracle and operator-ablation diagnostics. The index-level
excess is absent for a continuous annual mean and becomes pronounced as the
observable approaches concentrated hard-count geometry.}
\label{fig:ablation}
\end{figure}

The soft observable was
\[
S_y(\tau)=\sum_{m=1}^{12}
\operatorname{logit}^{-1}\left\{\frac{u_m-X_{y,m}}{\tau}\right\}.
\]
At $\lambda=4$, index-level rejection remained near the oracle/native range for
$\tau\in\{2,1,0.5,0.25,0.1\}$ and increased sharply near the hard-count
limit. A separately seeded near-zero follow-up with $n=180$ pooled
trajectories and $B=99$ extended the grid below 0.1.

\begin{table}[t]
\caption{Near-zero diagnostic for the concentrated profile. The last column is
the median index/oracle null-standard-deviation ratio.}
\label{tab:nearzero}
\centering
\small
\begin{tabular}{rrrrrr}
\toprule
$\tau$ & Oracle & Native & Index & Idx-only & SD$_I$/SD$_O$\\
\midrule
0.100 & 3.9\% & 2.2\% & 4.4\% & 2.8\% & 0.846\\
0.050 & 3.3\% & 2.2\% & 3.3\% & 1.7\% & 0.764\\
0.025 & 3.9\% & 2.8\% & 7.8\% & 5.0\% & 0.691\\
0.010 & 4.4\% & 3.9\% & 12.2\% & 8.3\% & 0.583\\
0.005 & 4.4\% & 3.9\% & 12.8\% & 8.9\% & 0.556\\
0.002 & 4.4\% & 3.9\% & 15.0\% & 11.1\% & 0.538\\
0.001 & 4.4\% & 4.4\% & 15.6\% & 11.1\% & 0.538\\
hard  & 3.9\% & 3.9\% & 16.1\% & 12.8\% & 0.532\\
\bottomrule
\end{tabular}
\normalsize
\end{table}

\begin{figure}[t]
\centering
\includegraphics[width=0.82\textwidth]{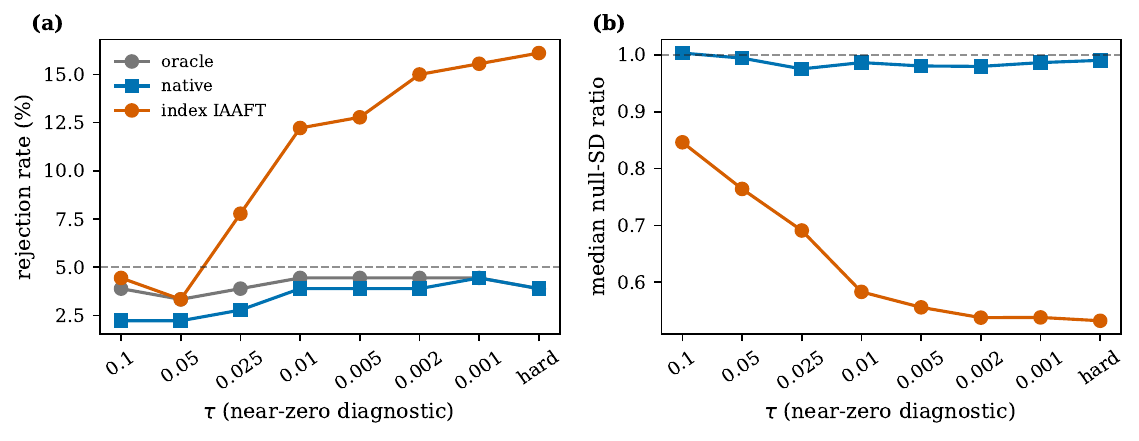}
\caption{Near-zero mechanism diagnostic. Index rejection and null contraction
increase progressively as the soft observable approaches the hard-count
geometry.}
\label{fig:nearzero}
\end{figure}

\section{Extremal-spacing, iteration, and phase diagnostics}\label{supp:spacing}

In two independently seeded small diagnostics (48 observed trajectories,
$B=99$ per trajectory), the hard-count annual IAAFT null regularized
upper-tail spacing relative to oracle and native nulls. At the hard limit, the
median ensemble mean number of adjacent exceedance pairs was 2.34 for index
IAAFT, 2.89 for native, and 2.88 for oracle. The median inter-exceedance-gap
standard deviation was 8.59 for index IAAFT and 9.95 for oracle. The mean
Ferro--Segers null was 0.972 for index IAAFT, 0.902 for native, and 0.892 for
oracle; corresponding null standard deviations were approximately 0.069,
0.132, and 0.132. At $\tau=0.1$ these differences were small.

The effect did not disappear with more IAAFT iterations. For the concentrated
hard-count condition, median normalized spectral error was 0.003189 after 10
iterations and 0.002859 after 30, 100, and 300 iterations. For
$\tau=0.01$, it was 0.001484 after 30 iterations and 0.001476 after 100 and
300. A global phase-dispersion diagnostic also did not support simple phase
locking: the median circular resultant of Fourier phases remained about
0.067--0.069 from the continuous annual mean through near-hard and hard-count
conditions.

\section{Within-fiber lumpability diagnostic}\label{supp:fiber}

For each Gaussian AR(1) trajectory $x$, a fiber mate $\widetilde x$ was
constructed by permuting exact amplitudes within
calendar-month $\times$ threshold-indicator class. Thus $x$ and
$\widetilde x$ had identical threshold indicators at every month, identical
annual count sequences, and identical calendar-month value multisets.
Generally, however, they had different standardized monthly Fourier-amplitude
targets.

Two independent native ensembles from $x$ supplied a same-$x$ Monte Carlo
baseline, and a third ensemble from $\widetilde x$ supplied the within-fiber
comparison. Each ensemble used $B=249$, with 90 fiber pairs for each of
$\lambda=0$ and $\lambda=4$.

\begin{table}[t]
\caption{Within-fiber distributional distances. Wasserstein distances are
scaled by the pooled standard deviation of the compared ensembles.}
\label{tab:fiber}
\centering
\small
\begin{tabular}{lrrrr}
\toprule
Projection & $\lambda$ & Same-$x$ & Fiber & Ratio\\
\midrule
Ferro--Segers & 0 & 0.086 & 0.175 & 1.93\\
Ferro--Segers & 4 & 0.090 & 0.335 & 3.31\\
Lag-1 correlation & 0 & 0.105 & 0.683 & 5.27\\
Lag-1 correlation & 4 & 0.108 & 0.535 & 4.55\\
Raw $T_{\rm diff}$ & 0 & 0.105 & 1.812 & 15.62\\
Raw $T_{\rm diff}$ & 4 & 0.114 & 1.867 & 15.28\\
Low-frequency share & 0 & 0.111 & 0.627 & 5.36\\
Low-frequency share & 4 & 0.099 & 0.379 & 3.97\\
\bottomrule
\end{tabular}
\normalsize
\end{table}

For Ferro--Segers, the fiber distance exceeded the same-$x$ baseline in 73.3\%
of pairs at $\lambda=0$ (one-sided sign-test
$p=5.45\times10^{-6}$) and 91.1\% at $\lambda=4$
($p=6.92\times10^{-17}$). KS comparisons rejected equality in 70.0\% of
$\lambda=4$ within-fiber Ferro--Segers pairs but only 1.1\% of same-$x$
controls. For raw $T_{\rm diff}$, all within-fiber pairs were distinguished by
the KS comparison for both seasonal profiles. Under $\lambda=4$, hidden
target-spectrum distance was strongly associated with transported-null
distance for $T_{\rm diff}$ (Spearman $\rho\approx0.90$).

\section{Additional SINCA details and sensitivities}\label{supp:sinca}

\subsection{Frozen cohort and protocol}

The 31-station cohort was selected using only station identity, coverage,
event-count, and gross scale-quality criteria. No station was added or removed
using a surrogate $p$-value, effect direction, null width, or visual inspection
of an inferential result. The fixed window was 2018-01-01 through 2025-12-28
(417 complete weeks). Daily coverage among retained stations was at least
97.36\%, and frozen monthly event geometry ranged from $\Neff=3.398$ to
5.673 (median 4.297).

The exact station list and the prospectively written protocol are
supplied as machine-readable supplementary files. The protocol fixes the
missingness rule, preprocessing design, native and index contracts, $B=499$,
station-level outputs, one-sided region sign-flip test, BY adjustment,
probability equalization, and all sensitivity analyses.

\subsection{Primary and equalized results}

Under the fixed 50 $\mu$g m$^{-3}$ threshold, the index contract
rejected at 31/31 stations and the native contract at 2/31, giving 29
index-only, zero native-only, two both, and zero neither. The one-sided exact
region sign-flip test over eight regions gave $p=0.00390625$. After BY
adjustment, 27 index station results and zero native station results remained
significant.

The median observed Ferro--Segers statistic was 0.535; median native and index
null centers were 0.557 and 0.993. The median center gap was 0.422. The median
index/native null-width ratio was 1.120. The prespecified correlation between
$\Neff$ and log width ratio was $-0.379$ with 95\% region-cluster bootstrap
interval $[-0.624,0.136]$, so the predicted width-contraction mechanism was
not supported.

Probability equalization moved realized $\Neff$ close to 12. Index rejection
fell to 6/31, native rejection to 1/31, and index-only decisions to 5/31.
The one-sided region sign-flip value became $0.0625$, and no station remained
significant after BY adjustment under either contract. The median center gap
fell to 0.097, decreased at every station, and the index Monte Carlo $p$-value
increased at 30 of 31 stations.

\begin{figure}[t]
\centering
\includegraphics[width=0.85\textwidth]{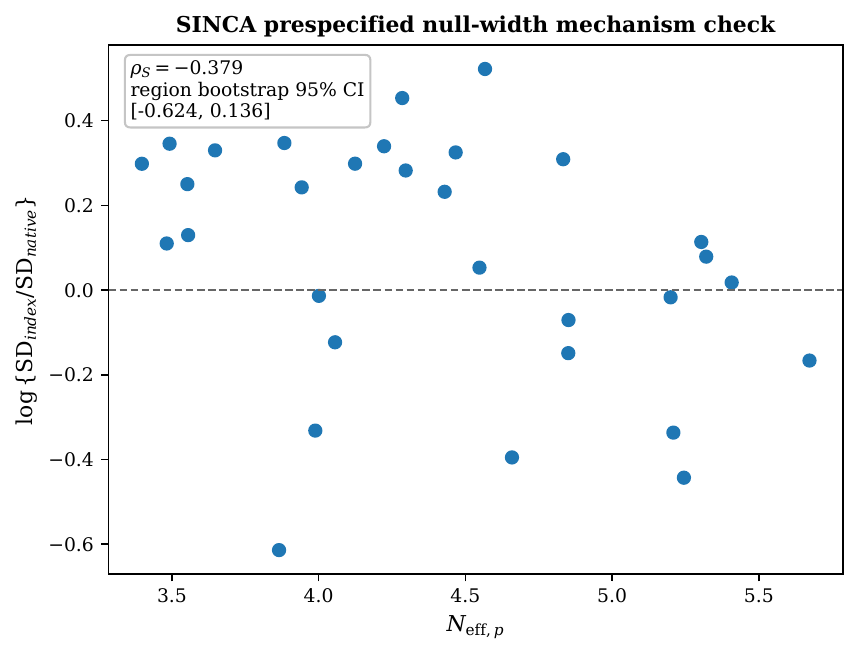}
\caption{SINCA width-mechanism diagnostic. The prespecified association
between seasonal concentration and null-width contraction is not supported,
which motivates characterizing the complete projected null distribution
rather than variance alone.}
\label{fig:sincawidth}
\end{figure}

\subsection{Sensitivity analyses}

With Ferro--Segers $q=0.85$, the fixed-threshold branch gave 31 index
rejections and 10 native rejections; the equalized branch gave two index and
zero native rejections. For normalized $T_{\rm diff}$, the fixed branch gave
four index and one native rejection, and the equalized branch five index and
two native rejections. These differences reinforce the data-processing
interpretation: the same kernel-level transport defect can project differently
onto different statistics.

The exposure-standardized missingness sensitivity was technically defined for
22 stations. It gave 22 index rejections and one native rejection, with 21
index-only and one joint rejection. Thus the primary applied separation was
not created by expected-value contributions on sparse missing days.

\section{Reproducibility files}\label{supp:repro}

The versioned archive supporting the manuscript is Zenodo v2.0.0,
DOI \href{https://doi.org/10.5281/zenodo.22713989}
{10.5281/zenodo.22713989}. It contains:

\begin{itemize}
\item the frozen SINCA cohort, inferential protocol, protocol manifest, and
analysis-ready 2018--2025 PM$_{2.5}$ input;
\item station-level fixed-threshold, equalization, and missingness
sensitivity results;
\item executable Python scripts for the SINCA analysis;
\item the confirmatory synthetic outputs and original probability
equalization evidence;
\item Gaussian, near-zero, iteration, phase, spacing, and fiber diagnostic
outputs and scripts;
\item integrity manifests and SHA-256 checksums.
\end{itemize}

The versioned archive is also supplied with the submission as a reproducibility supplement.

\end{document}